\documentclass[10pt,pre,twocolumn]{revtex4}
\usepackage{subfigure}
\usepackage{graphicx}
\usepackage[floatfix]{epsfig}
\usepackage{amsmath,amssymb,amsthm}
\usepackage[latin1]{inputenc}

\begin{document}

\title{What do we learn from correlations of local and global network
  properties?} 

\author{Magnus Jungsbluth, Bernd Burghardt and Alexander K. Hartmann} 

\affiliation{Institut f\"ur Theoretische Physik, Universit\"at
  G\"ottingen, Friedrich-Hund Platz 1, D-37077 G\"ottingen, Germany}

\date{\today}

\begin{abstract}
In complex networks a common task is to identify the most important or
``central'' nodes. There are several definitions, often called
centrality measures, which often lead to different results.
Here we study extensively correlations between four local and global 
measures namely
the degree, the shortest-path-betweenness, the random-walk betweenness
and the subgraph centrality on different
random-network models like Erd\H{o}s-Rényi, Small-World and
Barabási-Albert as well as on different real networks like metabolic 
pathways, social collaborations and computer 
networks. Correlations are quite different between the real networks
and the model networks questioning whether the models really reflect all
important properties of the real world.
\end{abstract}

\maketitle

\section{Introduction}
Theories for complex networks have attracted much attention in the
last few years. Started by the social sciences \cite{scott2000} the research 
incorporates disciplines ranging from the social sciences over
biology to physics.  There are studies for example on analytical
properties of certain network models, on attack vulnerability of real
networks \cite{Attack} 
or in the prediction of epidemics \cite{brockmann2006},
just to name a few. Extensive reviews are given in Refs. 
\onlinecite{albert2002,NewmanBasic}. 

The main focus has been on the so-called
\textit{scale-free} networks that have a degree distribution
$P(k)$ (the probability for a node to have k edges to other nodes) 
obeying a 
power-law. The exponent $\gamma$ of these power-laws $P(k) \propto k^{-\gamma}$
is typically close to 3 for many real networks. There have been some
attempts to explain this behavior based on the seminal work of Barabási
and Albert \cite{BA1, BA2} who explain the scaling behavior by a preferential
attachment mechanism during network growth. These models are based on the 
notion that the ``importance'' (often called {\em centrality}) 
of a node in some cases is given by the
number of connections of a node.
Nevertheless, it seems clear not all
properties of a complex real-world system can be explained by models based
on this ingenious yet simple mechanism.
Since the degree is a very local measure on a network it
is not necessarily the best choice to characterize all types of networks.
Over the years a few other measures \cite{betweenness,
  NewmanRDW, Subgraph} for the importance of nodes have been proposed that
actually measure global properties of the whole structure. 
In these publications, examples are shown where some nodes in a network
have a small degree, yet they play an important role for the network.
Hence, these more globale measures may deserve more attention.
Nevertheless, we are not aware
of a thorough comparison of these measures on different
model and real-world networks. Due to the relatively small number of studies
on these more complex measures, it is so far unclear wether they
are indeed better suited to identify important nodes in networks.

For a given network, the different measures may be strongly correlated, i.e.\
a node, which has a high importance found when measuring using 
one measure, appears also important
when using another measure, and vice-versa. If this was generally
true for all networks, then it would be sufficient to study just one measure. 
E.g., if all measures were strictly monotonic and simple functions
of the degree, then the degree would be indeed the key quantity to study.
If, on the other hand, 
different measures are not strictly correlated, then nodes, which
yield a high importance even for different measures, can be regarded indeed as
key nodes for a given network. Also it might be that there are 
nodes which obtain a high
value for one measure, but not for another measure.
In this case either one measure is not suitable for the description 
of a given network, or, if this is not systematically true, these 
nodes have to
be studied more closely, to understand a network's behavior.
In any case, it appears that studying the {\em correlations} of different
local and global
properties of nodes is a promising way to understand networks much better
than just to look at the distributions of single, maybe even solely local
properties. For this paper, we have systematically studied several local
and global network measures for different types of network models and for a
couple of
networks describing real-world data. As in some previous
studies, we find that the distributions
of single measures show in most cases the well-known scale-free
behavior, if the network shows scale-free behavior in the degree-distribution. Nevertheless, 
the standard network models are {\em not} capable
to reproduce in many cases the
complicated correlation signatures we find here in the real-world 
data. Hence, we propose that the systematic study of these correlations
as a much better tool to study networks and a comparison
of these correlations should be a suitable
criterion to evaluate the validity of network models.

This paper is organized as follows: In section \ref{sec:measures} we
introduce the centrality measures we have used in our studies. Section
\ref{sec:random} gives an overview of the random networks we have 
considered. In
section \ref{sec:real} we present the real networks that we have studied and
how they have been constructed. In section \ref{sec:results} we show our
results and in section \ref{sec:summary} we give an outlook to possible
future directions of research.

\section{Centrality measures}
\label{sec:measures}
In mathematical language a network (also often called graph) is a pair
 $G=(V, E)$ consisting of
a discrete
set of nodes (also called vertices) $V$
 and a discrete set of edges $E\subset V\times V$. 
We are
only interested in undirected networks and therefore an edge $e=\{i,j\}$
is a 2-set of nodes containing the two nodes $i,j$ connected by the 
edge. 
 A component of a network is a subset of nodes
with the following properties: Each node is reachable from each
other node via some path (i.e.\ a directly connected sequence) of edges 
and it is impossible to add another node
from $V$ without breaking the first requirement. In that sense it is a
maximal subset. A network may consist of more than one component but we
are mainly interested in those networks that consist of one component.
Networks with more that one component can be decomposed into
a set of smaller one-component networks. 
In the following $n=|V|$ denotes the number of nodes and $m=|E|$ denotes
the number of edges. We assume that there is an arbitrary but fixed
order on the set of nodes so you can enumerate them. Each node
has therefore a natural index. 

The most prominent centrality measure of a network is the so called
\textit{degree}, which is the number of edges incident to a node, i.e.\  
it's number of neighbors. It
can be calculated in
$O(1)$ if an appropriate network representation is used. 
The degree has been used very often to describe the importance of a node.
For example for computer networks, where the computers are represented by
nodes and the physical network connections by links,
routers and servers, which play a central role in these systems, 
are connected to many other computers.
Hence, networks are often
characterized by their degree-distribution. The class of scale-free
networks, that is networks with a power-law distribution, has been in
the focus of interest because many real-world networks reveal a
scale-free degree distribution. 

On the other hand, the degree is just a {\em local}
 measure of the centrality of
a node. For example in a motor-way network, where the nodes represent junctions
and the edges represent routes, there can be very important junctions, which
only connect few routes, but a breakdown of one of these junctions leads
to a major traffic congestion. Hence, other measures have been
introduced, which are intended to reflect to {\em global} importance
of the nodes for a network.

A measure of centrality 
that takes advantage of the global structure of a network is
the \textit{shortest-path betweenness} or simply \textit{betweenness}
of a node $i$, which is defined as the fraction of shortest paths
between all possible pairs of nodes of the network that pass through node
$i$. Let $g_i^{(st)}$ be the number of shortest-paths between node
$s$ and $t$ running through node $i$ and $n^{(st)}$ the total number of
shortest-paths between  $s$ and $t$. Then the \textit{betweenness}
$b_i$ for node $i$ is given by

\[
  b_i = \frac{2 \cdot \sum_{s<t}{g_i^{(st)} / n^{(st)}}}{n(n - 1)}
\]
The normalization $n(n - 1)/2=\sum_{s<t} 1$ 
ensures that the value of the betweenness
is between zero and one. 
This measure has been  introduced in social sciences (see \cite{betweenness} 
and \cite{betweenness2}) quite a while ago. The algorithm we use to calculate
the betweenness is presented in Ref.\ \onlinecite{NewmanSP} and has a 
time-complexity of $O(mn)$. That means this
algorithm can handle rather large networks really efficiently. 

The logical background of the betweenness is that the flow of information,
goods, etc., depending on the type of network, can be in some way directed
in a deterministic way. In particular the full network structure must be
known for each decision. Nevertheless, e.g.\ if all people decide to take the
same single shortest route to the center of a city, 
this might result in a large value of the overall travelling times.
Also, there may be networks,
e.g.\ social networks, the nodes representing persons and the edges
representing personal relations, where the information flow is not controlled
externally or deterministically and the full network structure is not
known to all players.
A recent proposal for the so called \textit{random-walk betweenness} 
(\textit{RDW betweenness}) by
Newman \cite{NewmanRDW} models the
fact that individual nodes do not ``know'' the whole structure of
the network and therefore a global optimum assumption is not very
convincing.  Within this approach, 
random walks through the
network are used as a basis for calculating the centrality for each
node: The random-walk betweenness of a node $j$ is the fraction of
random walks between node $s$ and node $t$ passing through $j$
averaged over all possible pairs of source node $s$ and target
node $t$. Loops within the random walks are excluded by using
probability-flows for calculating the actual
RDW betweenness. After a simple calculation \cite{NewmanRDW}
one arrives at an algorithm, which looks like as follows:

\begin{enumerate}
	\item Construct the adjacency matrix A and the degree matrix D
\begin{align}
A_{ij} &= 
\begin{cases} 
 1, & \text{iff edge \{i, j\} exists} \\ 
 0, & \text{else} \\ 
\end{cases} \nonumber
\\
D_{ij} &= 
\begin{cases} 
 k_j = \sum_{i'} A_{i'j}, & i = j \\ 
 0, & \text{else} \\ 
\end{cases} \nonumber
\end{align}
  \item Calculate the matrix D - A
  \item Remove the last row and column, so the matrix becomes
    invertible (any equation is redundant to the remaining ones)
  \item Invert the matrix, add a row and a column consisting of zeros
    and call the resulting matrix T
\end{enumerate}

Note that so far the calculated quantities do not depend on $i$ or $s,t$.
Now the random-walk betweenness $b_i$ for node $i$ can be calculated by 
\[
	b_i = \frac{2 \sum_{s < t} I_i^{st}}{n (n - 1)}
\]
where
\[
I_i^{st} = 
 \frac{1}{2} \sum_j A_{ij} \left|T_{is}-T_{it}-T_{js}+T_{jt}\right|  
\]
if $i \neq s$ and $i \neq t$ and $I_i^{st}$ equal to one if $i$ is equal to $s$
or $t$. Note that although the RDW betweenness is based on a random quantity,
its calculation is not at all random. Hence, any scatter observed in the
data is due to the networks structure not due to fluctuations of
the measurement.
It is possible to implement the calculation of the RDW betweenness with
time-complexity $O((m + n)n^2)$. The drawback is the considerable 
amount of computer memory needed
since this algorithm uses a adjacency matrix and other matrices of the
same dimension. Hence the memory consumption has the order
$O(n^2)$. Sparse-matrix methods could make the situation better since
most networks have sparse adjacency matrices but that would worsen the
time-complexity which is not desirable.

The fourth measure we use within this study is the
\textit{subgraph centrality} \cite{Subgraph} (\textit{SC}), 
which is based on the idea that the importance of a node should depend on its
participation in local closed walks where the
contribution gets the smaller the longer the closed walk is. The
number of closed walks of length k starting and ending on node $i$
in the network is given by the local spectral moments $\mu_k(i)$ of
the networks adjacency matrix $A$ which are defined as
\[
  \mu_k(i) = (A^k)_{ii}
\]
The definition of the \textit{SC} for node $i$ is then given by
\[
  C_S(i) = \sum_{k=0}^\infty \frac{\mu_k(i)}{k!}
\]
Albeit it is possible to directly calculate the series directly it
would not be overly efficient to do so. It is shown in \cite{Subgraph}
that it is possible to alternatively calculate the adjacency matrix's
eigenvalues $\lambda_i$ and an orthonormal base of eigenvectors $v_i$
for a network. Then the subgraph centrality $C_s$
for node $i$ can then be calculated via
\[
  C_S(i) = \sum_{j=0}^n [(v_j)_i]^2 e^{\lambda_j}\,.	
\]
This measure generally generates values with high order of
magnitude and is not in some way limited. We tried to normalize
with $C_s(1)$ of a fully connected graph with the same number of
vertices (all vertices are equal so every vertex has the same subgraph 
centrality), but this gave us values beyond machine precision for graphs
larger than 5000 vertices, i.e.\ even much larger than the values we observed
for the networks under consideration.
 Hence, we used the non-normalized values. 

\section{Random-Network Models}
\label{sec:random}
We compared the different measures on different random-network models,
namely the Erd\H{o}s-Rényi (ER) model \cite{ER1,ER2,ER3}, the
Small-World (SW) model \cite{SW1, SW2, SW3} and the
Barabási-Albert (BA) 
model \cite{BA1, BA2}. The ER model consists of random networks of
a fixed number of nodes $n$ and for each pair of nodes an edge is
added with probability $p$. The degree distribution of this model is 
Poissonian.

The SW model is also characterized by a fixed number of nodes $n$, 
but here the nodes a placed on a regular grid. An instance is generated in
two steps. First, each node is
connected to its $k$ nearest neighbors.  In the second step, 
each edge is reconnected to one
random node with probability $p$ (i.e.\ the other node remains). 
Most SW networks studied are based on a one-dimensional grid with
periodic boundary conditions, i.e.\ the nodes are
ordered on a circle. The degree distributions of these networks
interpolates between a delta peak at $k$ for $p=0$ and the Poissonian
distribution for $p\to1$.

The BA model is  the only growth model studied here. In this case the networks
are created by a
so called preferential attachment mechanism. Each generated random network
 starts with $m$ nodes and new
nodes are added consecutively, one after the other. 
A new node is immediately connected to
exactly $m$ of the already existing  nodes, which 
are chosen randomly. The higher the degree $k$ of an existing
node the bigger is the chance that it is selected as neighbor. 
Hence, the probability
for a node $i$ to get selected is given by its degree $k_i$ 
divided by the sum $\sum_j k_j$ of all
degrees of all currently existing nodes of the network. To efficiently
generate these networks we used a list, where each node $i$ is contained
$k_i$-times. For each newly added node we select $m$
different elements randomly from the list and connected them to the new
node. The resulting degree distribution follows a power-law with exponent 
$\gamma=3$ \cite{Bollobas} in the limit of large degrees (in the tail
of the distribution).

It is also possible to get different exponents in the tail by adding a
certain offset $k_0$ to the probability of selecting a certain vertex,
so the total probability goes as $k + k_0$. This yields an exponent of
$\gamma = 3 + k_0 / m$ \cite{NewmanBasic} in the tail of the distribution. $k_0$ may be explicitly
negative as long as it is $-m < k_0 < \infty$.

For all random-networks we prohibited parallel edges between two
nodes and self-loops, i.e.\ for the BA model, each node $i$ can be
selected from the list only once. Additionally we extracted the largest
component for the ER networks and the SW networks. Note that the
BA model generates fully connected networks. 

\section{Real Networks}
\label{sec:real}

It is well known that the models presented in the last section are able
to reproduce some of the characteristics of real-world networks. The most
realistic models for many applications are the BA model and related
models based on growth mechanisms, which reproduce the power-law
behavior of the distributions of the degree and some other
centrality measures \cite{albert2002,NewmanBasic}. 
As indicated above, we propose in this paper to go beyond
measuring distributions of local or global properties, by considering 
correlations between different measures. Hence, to investigate whether
these most common models are also able to reproduce these complex
characteristics of real-world networks, we have to compare with the results
of at least some real-world networks.

We took data from publically available databases, as given below.
 In all cases, we treated the network as undirected,
unweighted network. This in some cases not a good model but to
examine all the networks in exactly the same way, we have chosen to do
so. In all cases, where the networks consisted of more than one
component,
 we only used the largest component of the network,
since especially the random-walk betweenness is not defined properly
on a network having more than one component. Additionally we eliminated
all self-loops (edges connecting the same node) and parallel edges in the real-world networks,
if present. 

We have studied the following five networks.
\begin{itemize}

\item {\bf Protein-protein interaction in
Yeast (PIN)\quad } The data was obtained from the COSIN database 
\cite{cosin}. In the PIN network each node represents a certain 
protein and an edge
is placed between them if there has been an observation of an interaction between the two proteins 
in one of various experiments.

\item  {\bf Metabolic pathways \cite{MetabolicPath}
of the E. Coli bacteria (ECOLI)\quad}
The ECOLI network was obtained by using the API of the
KEGG \cite{kegg} database plus using the file "`reaction.lst"' from
the KEGG LIGAND database. The latter is needed to separate the educts
and the products of a reaction, since the API only outputs which
compounds are involved in the reaction. All compounds that are
catalyzed in any way by enzymes of the E. Coli are used as nodes and an
edge is placed between two nodes if there exists a reaction which
has one compound on one side of the reaction and the other compound on
the other side. 

\item {\bf Collaboration network of people working in 
computational geometry (GEOM)\quad}
In the GEOM network obtained from Ref.\ \onlinecite{geom}, 
each node represents 
an author from
the Computational Geometry Database with an edge between two authors
if they wrote an article together.

\item {\bf Network of autonomous systems (AS)\quad}
The AS network
is a computer network extracted via trace routes from the Internet
containing routers as nodes and real-world connections between them as
edges
(in fact virtual connections since the router's known hosts table
determines which nodes can be reached from a given point in the
network).  The data for AS was obtained also from the COSIN 
database \cite{cosin}.

\item {\bf Network of actors
collaboration (ACTORS)\quad}

The data was obtained from the Internet Movie Data Base \cite{imdb}.
Nodes represent actors. Since the database is very huge, we restricted
our study to films from the UK after 2002. Nodes are connected
by an edge, if the corresponding actors appear in the same film.

Unfortunately, the ACTORS network did not yield meaningful results because the
underlying data was quite ``noisy'': In movies with a lot of actors
listed in the data base, 
even the less important parts get a high connectivity. 
Thus, we observed for all measures given above 
a large scatter of the data points and very small correlations
between them.
Furthermore, we doubt that defining a network of actors 
in this way is meaningful,
because usually it is not the actors who decide with whom they interact 
in a film, but the producers who select the actors.
Therefore we do not show here any plots for  this network type.

\end{itemize}

Note that all networks created in this way are of size 
less than 10.000 nodes, which allows to  compute the measures defined in Sec. 
\ref{sec:measures} easily.

\section{Results}
\label{sec:results}
For all random models we have used a graph size of $n=2000$ nodes and
drew $100$ representatives from the ensemble of possible networks. After
calculating the four different measures for each network, we averaged
over all representatives to get smooth distributions for each measure
and network-type. For the real networks we just calculated all measures for
each given network, clearly no average can be performed here.

Since we consider four types of measures we can calculate 6 types of 
measure1-measure2 correlation
plots for each graph model and each real-world graph. Since we have studied the
three different graph ensembles for several values of the 
parameters, e.g.\ for the edge probability $p$, this is
 totalling in several hundred possible plots. Many of these
plots show strong correlations between the two quantities considered 
and give no qualitative information
beyond that. Hence, we restrict ourselves here to the most interesting cases,
which keeps also the length of the paper reasonable. 

\begin{figure}[htb]
\includegraphics[width=\columnwidth]
{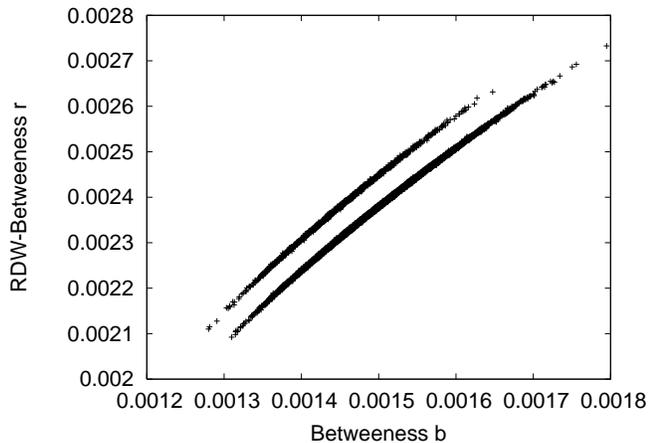}
\caption{Correlation of betweenness and RDW betweenness for 100 ER-networks
  with $p = 0.10$ and $n = 2000$.
\label{fig:er_corr_betweenness_rdw}}
\end{figure}
All Erd\H{o}s-Rényi networks, where we performed the analysis always for the
largest component, show high almost linear correlations
between any two measures (not shown) for all probabilities $p \in
{0.05,0.10, 0.15}$
 we have investigated.
 The data points of any measure1-measure2 correlation
plot lie on the data points of the averaged ensemble. This seems to
indicate that indeed different measures are equivalent to each other
and that, in order to characterize how important different nodes are,
 it might be sufficient to look at the degree, which is a local
quantity and simple to calculate. Note that the ER model is the most
simple model considered here, and below we will find examples, in particular
for real networks, which exhibit a much more complex behavior. Nevertheless,
even the ER networks show a behavior in one case, which appears to be very
strange.
We observe some sort of clustering in the
the correlation-plot of betweenness against RDW betweenness as can be
seen in the scatter plot over all instances in 
Fig.\ \ref{fig:er_corr_betweenness_rdw} for the
large edge probability $p=0.1$. It seems that
there are essentially two types of nodes belonging to two 
different correlation functions. Note that this splitting into two
different behaviors is more dominant the higher the edge-probability
$p$ of the generated networks is, i.e.\ the more likely it is that
each graph of the ensemble consists only of one component.
In particular for graphs with
small average degrees up to 50, this behavior is hardly visible.
So far, we do not
understand this kind of symmetry breaking. Since the two measures are
identical on star-networks and the random walk betweenness 
generally gives higher
higher scores for nodes that lie slightly off shortest paths in the
network, such local irregularities might be an explanation for this
behavior.

\begin{figure}[htb]
\includegraphics[angle=270, width=\columnwidth]{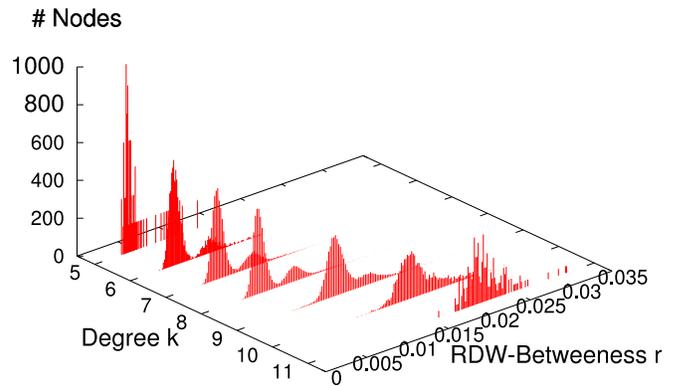}
\caption{Correlation of degree and RDW betweenness for 100 SW-networks
  with $p = 0.05$, $8$ nearest neighbors and $n = 2000$ with corresponding 
RDW betweenness distribution.
\label{fig:sw_3d_distribution}}
\end{figure}

For the  Small-World model we have studied values $k=8,16,24$
and $p=0.05,0.10,0.15$. We observe usually
 moderately high correlations, but lower than for the ER model (not shown 
directly, see below).
For the degree-RDW betweenness correlation, the data points are not uniformly
distributed, similar to the betweenness-RDW betweenness we have shown for
the ER graphs above. This can be seen here better when
looking at the correlation using a three-dimensional plot of impulses, 
rather than a scatter plot, see
Fig.\ \ref{fig:sw_3d_distribution}.
The "`oscillation"' that can be observed is also consistently
present in the overall distribution of the RDW betweenness (averaged over 
all degrees). Hence, here also two types of vertices seem to be present, but
the distinction is weaker than above. Even for very small
probabilities (i.e.\ $p = 0.0001$, $n = 2000$, $k=8$) the two peaks are
visible though they are very close together. The gap between the two
peaks gets larger the higher the rewiring probability $p$. Here, the
difference seems to be strongly related to the re-wiring of the nodes, 
because for the case $k=8$, i.e.\ the degree of the corresponding $p=0$
network, the two peaks in the distributions are most clearly
separated.
\begin{figure}[htb]
\includegraphics[width=\columnwidth]{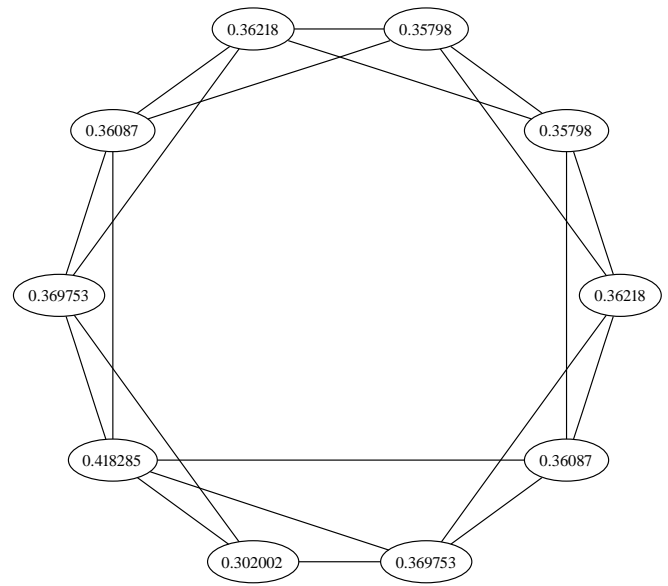}
\caption{A SW-network with ten nodes. 
The numbers indicate the random-walk betweenness of a node. The node
with a newly gained edge by rewiring obtains a much higher value than the
others, while the node which lost an edge obtains a much lower value.} 
\label{fig:sw10}
\end{figure}
Even for very small network sizes like 20 nodes, it is possible to see
two different peaks. Consider for example the 20 node network shown
in Fig. \ref{fig:sw10}, where just one edge has been rewired. The
node with the highest value for the random-walk betweenness is the one
that gained an additional edge by re-wiring. It is also visible that
nodes with the same degree get different values for the
RDW-Betweenness, which become smaller with growing distance to the most
important node. This explains why the peaks get smeared out: It is
because the nodes that get new edges influence those nodes that stay
the same from a degree point of view. In general even nodes that keep
their degree constant but gain crosslinks to high RDW betweenness
nodes get a similarly high RDW betweenness. So one explanation of the
peaks would be that the lower peak is a smeared out version of the
one-value peak before rewiring and the the peak for higher
RDW betweenness values appears because rewired nodes get a much higher
RDW betweenness.  

\begin{figure}[htb]
\includegraphics[width=\columnwidth]{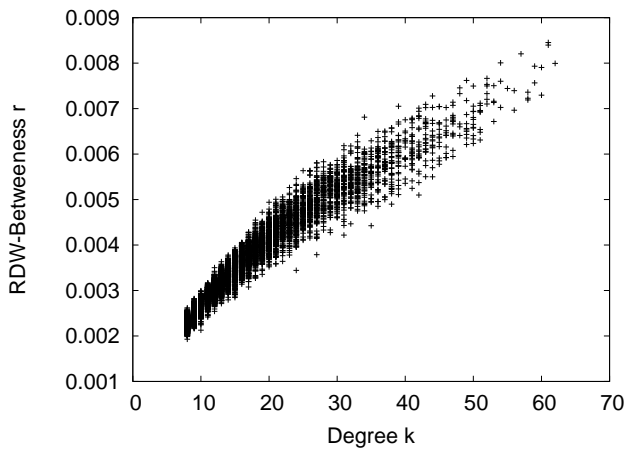}
\caption{Correlation of degree and RDW betweenness for 100 BA networks
  with $m = 8$ and $n = 2000$.
\label{fig:all_corr_degree_randomwalk}}
\end{figure}

The Barabási-Albert networks,
where we have studied values $m \in {8, 16, 24}$,
 show again almost linear correlations for all
combinations of measures. Nevertheless, the correlations were not as clear
as for the Erd\H{o}-Rényi networks, i.e.\ we observed a much larger
scattering of the data, but in the same order of magnitude as for the SW graphs. 
An example can be seen in
Fig.\ \ref{fig:all_corr_degree_randomwalk}. Here, we did {\em not} observe
any particular strange correlation for any combination of
measures, in contrast to the other two models.

Hence, to summarize the study of the correlations for the random graphs 
(results for the distributions, in particular exponents in case of
power-law behavior, see below), we find most of the time a strong correlation
between {\em all} different centrality measures, hence the degree
is almost sufficient to characterize the importance of a node. This
statement is certainly not true for many networks based on real-world
data, as we will see next.

\begin{figure}[htb]
\includegraphics[width=\columnwidth]{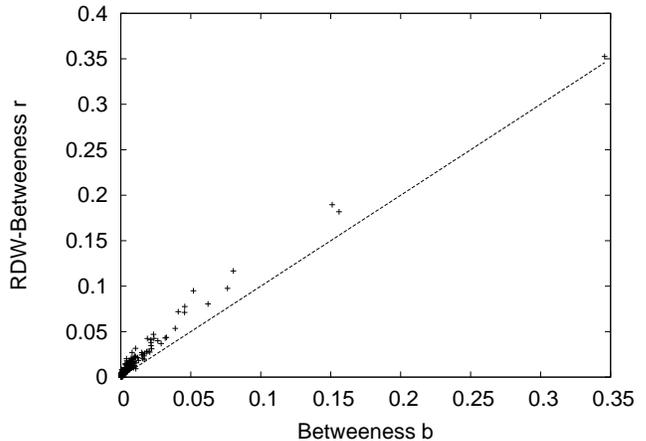}
\caption{Correlation of betweenness and RDW betweenness for the AS network.
\label{fig:AS_corr_betweenness_randomwalk}}
\end{figure}

The AS network exhibits a positive correlation for all combinations of
measures. Nevertheless, the results show some aspects of the
behavior which is strongly different from the
networks models discussed previously. 
E.g.\, a scatter plot of betweenness
against RDW betweenness is shown in 
Fig.\ \ref{fig:AS_corr_betweenness_randomwalk}. One can see that 
the scatter of the data points appears is always very small.
This indicates that the fluctuations
generated by the local structure around a node are always of the same order
of magnitude, irrespectively of the absolute value of a quantity. Furthermore,
even more interestingly, we observe that almost all data points
obey the inequality $r\ge b$.
So far we do not have an explanation for this effect,
which is not present in the data for the network models.

\begin{figure}[htb]
\subfigure[The PIN-network.\label{fig:yeast_corr_degree_subgraphcentrality}]
{
\includegraphics[width=\columnwidth]{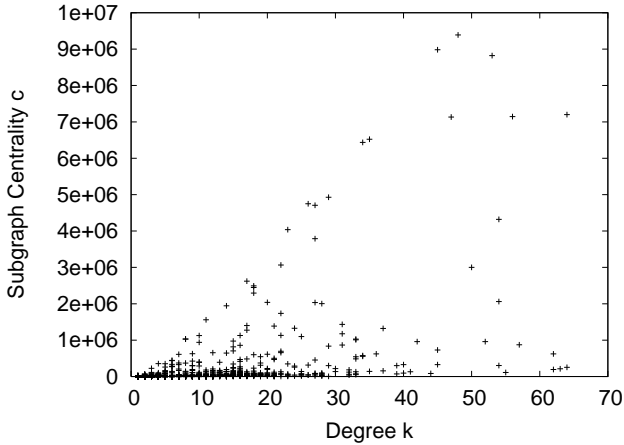}
}
\subfigure[A BA network with similar degree-distribution.\label{fig:ba_yeastcomparison_corr}]
{
\includegraphics[width=\columnwidth]{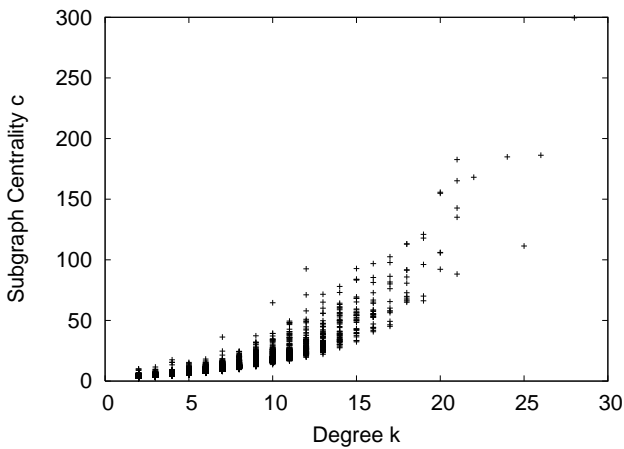}
}
\caption{Comparison of degree-SC correlation between BA-Model and PIN-network}
\end{figure}

The calculations on the PIN network presented high correlations on all
combinations of measures (not shown) except the degree-SC correlation plot,
see Fig.\ \ref{fig:yeast_corr_degree_subgraphcentrality}. Here 
you can see two "`branches"' that contain the data points. 
Thus, there are two types of vertices. For one type, the number of
closed walks increases exponentially with the degree. This is the behavior,
we have for complete (sub-) graphs (cliques), 
i.e.\ where each protein interacts with each other member of the (sub-)
graph. On the other hand, there are proteins, where the participation
in closed walks does not increase at all with the degree, which means that
these proteins, although possibly with a large number of interacting
partners, participate
nevertheless only loosely in the overall interaction network. Note
that for large degrees, there seem to be even proteins, which
interpolate between the two limiting behaviors.
This behavior is quite the different to what you find with for example 
for the BA networks
and it is a hint that the structure of this network cannot be modeled
with BA networks although it degree distribution, which we
have measured as well (see below), is still scale-free. To illustrate
this we tried to fit a BA network's degree distribution as good as
possible to the degree distribution of the PIN network and found the
best fit for $m = 2$ and $k_0 = -1$, although the BA networks for
these parameters have generally a smaller maximum degree than the PIN
network. As you can see in Fig.\ \ref{fig:ba_yeastcomparison_corr} the
correlation plots look completely different. For different values of
$m$ and $k_0$ the scales of the axes change (especially the SC yields
much higher values in the same order of magnitude as the PIN network)
but the generally behavior is consistent. 
A model for such an interaction 
network would have to take the existence of two types of proteins into account,
resulting in two different rules for the creation of the nodes.
In a recent study of the PIN network \cite{pinsc} which also uses a few centrality
measures it is found that high subgraph centrality is a better
hint for essential proteins than for example the degree. Thus it
fits nicely to our result 
that the degree and subgraph centrality are not strongly
correlated in this case.

\begin{figure}[htb]
\includegraphics[width=\columnwidth]{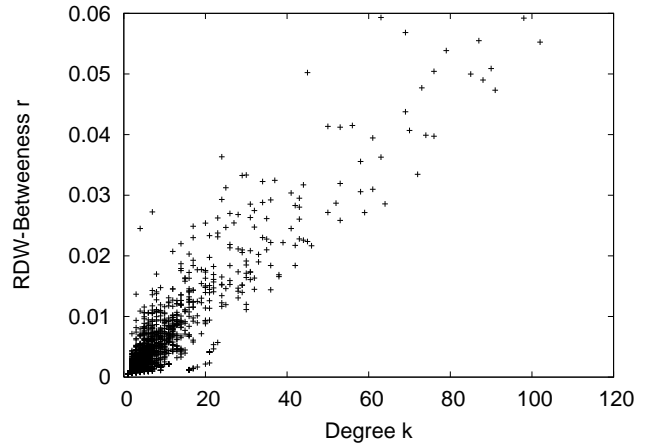}
\caption{Correlation of degree and RDW betweenness for the GEOM network.
\label{fig:geom_corr_degree_randomwalk}}
\end{figure}

\begin{figure}[htb]
\includegraphics[width=\columnwidth]{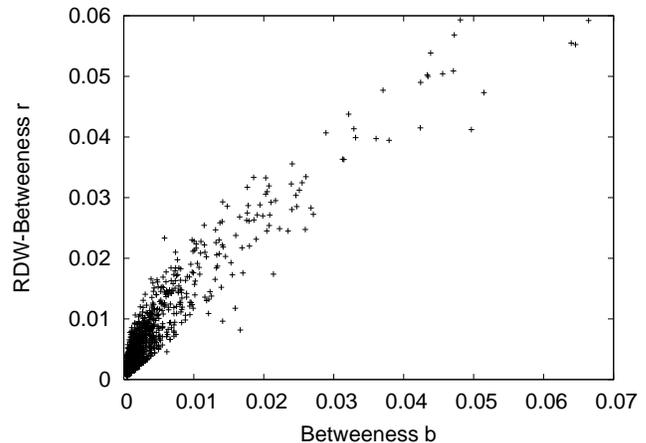}
\caption{Correlation of betweenness and RDW betweenness for the GEOM network.
\label{fig:geom_corr_betweenness_randomwalk}}
\end{figure}

For the GEOM network the measure1-measure2 correlation plots
show a quite scattered behaviour, i.e.\ much smaller correlations
than seen in the network models, see e.g.\ 
Fig.\ \ref{fig:geom_corr_degree_randomwalk}. Here we also 
 observe the $r\ge b$ feature in the
betweenness-RDW betweenness correlation plot, see 
Fig.\ \ref{fig:geom_corr_betweenness_randomwalk}. but the effect is 
even stronger in comparison to Fig.\ \ref{fig:AS_corr_betweenness_randomwalk}.
Hence, this inequality might be a property seen in many networks based
on real-world data and it certainly deserves a more thorough
investigation.

\begin{figure}[htb]
\includegraphics[width=\columnwidth]{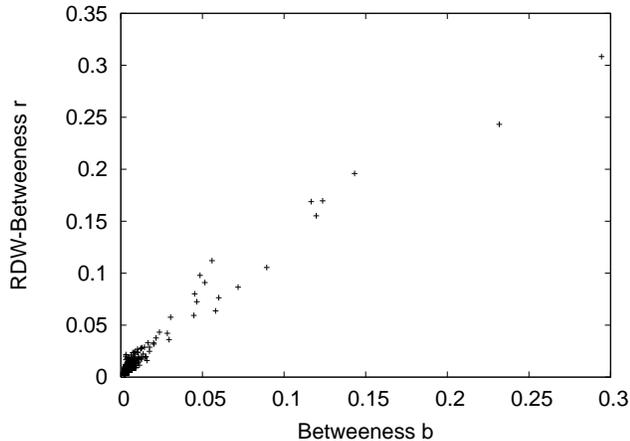}
\caption{Correlations of betweenness and RDW betweenness on the ECOLI-network}
\label{fig:pathways_corr_betweenness_randomwalk}
\end{figure} 

For the ECOLI network, the correlations of the 
measures range from moderately correlated highly to highly correlated, 
see e.g.\ Fig.\ \ref{fig:pathways_corr_betweenness_randomwalk}. 
In principle, the
plots look quite similar to those of the AS network.
Here we could
not observe any particular new properties, hence we do not go into 
further details for this network type. 

\begin{table}[htbp]
  \centering
  \begin{tabular}{l|c|c|c}
    Name & Degree &  Betweenness & RDW-Betweenness\\
    \hline
    AS & 1.54(4) & 1.66(3) & 1.55(3)  \\
    GEOM & 2.34(6) & 1.86(5) & 1.51(4) \\
    ECOLI & 2.87(9) & 2.18(9) & 3.1(1) \\
    PIN & 1.65(4)  & 1.82(4) & 1.66(3)\\
  \end{tabular}
  \caption{Power-Law Exponents}
  \label{tab:exponents}
\end{table}

\begin{figure}[htb]
\includegraphics[width=\columnwidth]
{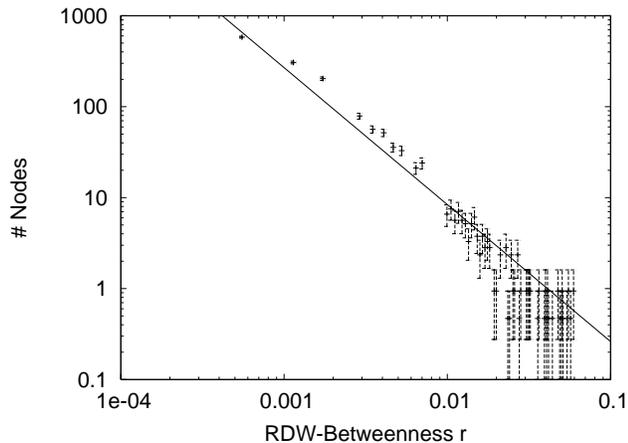}
\caption{RDW betweenness distribution for the GEOM network.
\label{fig:dist_geom_randomwalk.txt.eps}}
\end{figure}

Finally, we look just at the distribution of the centrality measures for
the real-world networks.
We find that all the real networks show a scale-free behavior,
see e.g.\ Fig.\ \ref{fig:dist_geom_randomwalk.txt.eps}.
We have fitted power laws $P(x) \propto x^\gamma$ to all data except for the 
the subgraph centrality, where the data
was distributed only over a small interval, so a fit would be
meaningless.  The power-law exponents we calculated can be found in
table \ref{tab:exponents}. This shows that when just looking at the
distributions of centrality measures, the behavior of the real-world
network is also found for the BA model. Goh et al. \cite{Goh} also found that
for this model the betweeness distribution follows a power law. Only when considering
correlations between different measures, one realizes that the so-far existing
models, although having provided much value insight, have to be extended 
and/ or modified, to really capture the behavior found in the 
behavior of proteins, metabolic pathways, humans and other systems
represented by networks.

\section{Conclusion}
\label{sec:summary}

In this paper we have studied four different local and global centrality
measures to analyze the behavior of different model and real-world
networks.
First, the choice which measure is ``most suitable'' depends on the network
that is used and which kind of information shall be extracted by calculating
that measure. There does not seem to be an overall best measure that is
optimal for all applications. 
The  shortest-path betweenness might be feasible if the
network can be assumed to contain global knowledge of optimal
routes. But even in this case, when much traffic is on the network,
it is certainly very often advisable to use non-shortest paths
to reach the destination as quick as possible.
In cases where participation in social sub-groups is of
interest the subgraph centrality might be best whereas in situations
where each node only passes information randomly to its nearest
neighbors the random-walk betweenness should be the method of
choice. 

Nevertheless, in order to understand really how a network is organized,
it sees not to be sufficient to study just one measure and
its distribution. We have seen that for all real-world networks
considered here, the distributions of all measures is indeed well
described by  power laws.
But when considering correlations between different centrality measures
we see that the most common 
 random network models reflect the truth only partially since
the scatter plots do look quite different compared to the real
networks.

It seems that network models have to be more specifically 
for each application. One single mechanism like preferential attachment, 
at least if being used as the only mechanism to create the graph,
is too simple to explain the complex properties of real-world networks.
Models that incorporate evolution and growth of
networks as represented for example in \cite{NetworkEvo} might be the
key to give deeper insight why many networks show a scale-free
behavior  for one of their properties and still differ from simpler
models like the BA model. 
Since each application will need its specific mechanism to generate
a network, proposing new models for specific applications 
is beyond the scope of this work.

\begin{acknowledgements}
The authors have obtained financial support from the
{\em VolkswagenStiftung} (Germany) within the program
``Nachwuchsgruppen an Universit\"aten'',
 and from the European Community
via the DYGLAGEMEM  program.
\end{acknowledgements}

\end{document}